\documentclass[
showpacs,
 amsmath,amssymb,amsfonts,
 aps,
 pra,
notitlepage
]{revtex4-1}

\usepackage[english]{babel}

\usepackage{graphicx}% Include figure files
\usepackage{dcolumn}% Align table columns on decimal point
\usepackage{bbm}% bold math
\usepackage{color}
\usepackage{enumerate}

\newcommand{\bra}[1]{{\left\langle{#1}\right\vert}}
\newcommand{\ket}[1]{{\left\vert{#1}\right\rangle}}
\newcommand{\bracket}[2]{\left\langle #1| #2 \right\rangle}

\def\ds{\displaystyle}
\def\Tr{\textnormal{Tr}}

\newtheorem{theorem}{Theorem}
\newtheorem{protocol}[theorem]{Protocol}

\usepackage{color}

\def\makenewenum#1#2{%
\newcounter{cnt#1}
\newenvironment{#1}%
{\begin{list}{\makebox[0pt][r]{#2}}%
{\setlength{\itemsep}{0pt}%
 \setlength{\parsep}{.2em}%
 \setlength{\leftmargin}{2.5em}%
 \setlength{\labelwidth}{.2em}%
 \usecounter{cnt#1}}}%
{\end{list}}}

\makenewenum{senum}{\bf{Step~\arabic{cntsenum}.}}

\begin{document}

\preprint{APS/123-QED}

\title{Quantum key distribution by phase flipping of coherent states of light}% Force line breaks with \\

\author{G. A. Barbosa$^{1}$}
\email{geraldoabarbosa@gmail.com}

\author{J. van de  Graaf$^{2}$}%
\email{jvdg@dcc.ufmg.br}

\author{P. Mateus$^{3,4}$}%
\email{pmat@math.tecnico.ulisboa.pt}

\author{N. Paunkovi\'c$^{3,4}$}%
\email{npaunkov@math.tecnico.ulisboa.pt}

\affiliation{
$^1$ QuantaSEC -- Consulting, Projects and Research in Physical Cryptography Ltd., Brazil
% Av. Portugal 158, Belo Horizonte MG 31550-000 Brazil
\\
$^2$ Departamento de Ci\^encia da Computa\c{c}\~ao, Universidade Federal de Minas Gerais, Brazil
\\
$^3$ Instituto de Telecomunica\c{c}\~oes, %-- Avenida Rovisco Pais 1, 1049-001 Lisbon, 
Portugal
\\
$^4$ Departamento de Matem\'atica, Instituto Superior T\'ecnico, Universidade de Lisboa %-- Avenida Rovisco Pais 1, 1049-001 Lisbon, Portugal
}

%\affiliation{
%\affiliation{%
\date{\today}% It is always \today, today,

\begin{abstract}
In this paper we present quantum key distribution protocol that, instead of single qubits, uses mesoscopic coherent states of light $\ket\alpha$ to encode bit values of a randomly generated key. Given the reference value $\alpha\in\mathbb C$, and a  string of phase rotations each randomly taken from a set of $2M$ equidistant phases, Alice prepares a quantum state given by a product of coherent states of light, such that a complex phase of each pulse is rotated by the corresponding phase rotation. The encoding of $i$-th bit of the key $r=r_1 \dots r_\ell$ is done by further performing phase rotation $r_i \pi$ (with $r_i = 0,1$) on the $i$-th coherent state pulse. In order to protect the protocol against the man-in-the-middle attack, we introduce a verification procedure, and analyse the protocol's security using the Holevo bound. We also analyse the possibility of beam splitting-like and of collective attacks, showing the impossibility of the former and, in the case of our protocol, the inadequacy of the latter. While we cannot prove full perfect security against the most general attacks allowed by the laws of quantum mechanics, our protocol achieves faster quantum key distribution, over larger distances and with lower costs, than the single-photon counterparts, maintaining at least practical security against the current and the near future technologies.
\end{abstract}

\pacs{03.67.-a, 03.67.Ac, 03.67.Dd}
\keywords{Quantum Key Distribution \and Coherent States \and Continuous Variables}

\maketitle

\section{Introduction}
\label{sec:introduction}

Quantum mechanics offers advantages when implementing data processing tasks in comparison to their classical counterparts. Arguably the most prominent one is the famous Shor algorithm for factoring numbers~\cite{sho:97}. While a practical implementation of a working scalable quantum computer, despite considerable success in the last decade, is still out of reach of today's technology, quantum cryptographic systems can already be bought on the market today. Quantum cryptography started with an early work from 1969 by Wiesner, who introduced notions of quantum multiplexing and quantum money, though he only managed to publish his work more than a decade later, in 1983~\cite{wie:83}. Based on his ideas, Bennett and Brassard introduced their famous four-state BB84 protocol for key distribution~\cite{ben:bra:84}. The unconditional security of quantum key distribution (QKD)~\cite{Lo:cha:99,sho:pre:00,may:01,sca:bsc:csr:dus:lut:pee:09} is a consequence of the laws of physics, and as such is stronger than the computational security of classical counterparts, based on unproven mathematical conjectures.

The physical systems that encode the bit values in the BB84 QKD protocol are qubits -- two-level quantum systems. So far, implementations of QKD protocols have (predominantly) used quantum optical systems. Thus, in the majority of such applications qubit states were naturally encoded in single-photon states (usually in polarisation). The use of single photons as carriers of qubits achieves theoretical perfect security, but has, nevertheless, a few drawbacks regarding single-photon detectors which are, at the present stage of technology: (i) relatively expensive, and (ii) too slow to facilitate the amount of information exchanged by today's average consumers. At the current technology stage commercial telecommunication detectors may operate at rates of $40$GHz (e.g., ref.~\cite{exp1}) or above while commercial single-photon detectors are still struggling below $50$MHz (see for example~\cite{exp2}) due to the need to quench the avalanche of a high numbers of electrons. Even laboratory superconducting nanowire detector devices operate at rates below 1 GHz as they need time for thermal dissipation after each excitation. Radical changes may appear in the technological horizon but this is way beyond the subject of this paper.

To meet such requirements, various different protocols for distributing keys using coherent states of light were studies in~\cite{ral:99,gro:gra:02,bar:03,bra:loo:05,qi:hua:qia:lo:07,jou:kun:lev:gra:dia:13,bar:gra:15,ott:man:pir:15,hua:hua:lin:zen:16,zhu:zha:dov:won:sha:16}. While the protocols based on coherent states  do achieve levels of security of single-photon QKD~\cite{lev:17}, to obtain the optimal key rates, one requires  low average photon numbers~\cite{lod_etal:07}, i.e., they too suffer from the above mentioned weaknesses. The use of mesoscopic coherent states indeed solves the above two problems (i) and (ii): multi-photon detectors (for average numbers of $10^2 - 10^4$ photons per pulse) need to be much less sensitive, and are thus less expensive, and can count  many more pulses per unit time.  %but the status of the protocol's security is still an unsolved issue. 
Instead of quadrature measurements the protocol presented in this paper only utilises phase encoding and detection in M-ry levels according to~\cite{bar:03,bar:gra:15}. In addition to that, the protocol~\cite{bar:03,bar:gra:15} uses a finite pre-shared secret key, constantly updating along its execution the shared randomness to achieve secure information transfer.

Recently, a secure public key encryption scheme based on single-qubit rotations was presented in~\cite{nik:08} and subsequently analysed in~\cite{sey:nik:alb:12} (see also a recent scheme~\cite{vla:rod:mat:pau:sou:15} based on quantum walks). Nevertheless, its standard optical applications using single-photon polarisation as a realisation of a qubit suffers from the same deficiencies as the above mentioned realisations of QKD. In this paper, based on the ideas of key distribution with continuous variables~\cite{bar:03,bar:gra:15}, and secure message transfer with single qubits~\cite{nik:08}, we present a version of a key distribution scheme in which bit values are encoded in (multi-photon) coherent states of light. Note that, unlike the protocol presented in~\cite{nik:08} which uses (single-photon) qubits, in our protocol states which encode single bit values are now from an infinitely-dimensional Hilbert space, and could thus, in principle, carry an unlimited amount of classical information. This makes the argument for the protocol's security, based on the Holevo theorem, rather non-trivial in our case (for the proof of the Holevo theorem, see for example~\cite{nie:chu:04}, Section 12.1.1, and the references therein). The main result of our paper is that such argument is indeed satisfied even for the particular infinitely-dimensional quantum states: given an ensemble $\hat\rho$ of particular coherent states used by Alice in our protocol, the amount of information carried by a single pulse is finite, bounded from above by the finite value of von Neumann entropy $S(\hat\rho)$; this way, the protocol is secure against Eve's attempt to learn Alice's choice of bases $k$, and thus subsequently read out Bob's encrypted key (for details, see Section~\ref{sec:security}). Note that once authentication for the users is established, the system does not demand a pre-sharing of keys to start the distribution stage, in contrast with~\cite{bar:03} or~\cite{bar:gra:15}. Therefore,  no courier is ever needed to refresh keys.  Coherent states with mesoscopic number of photons are much easier to construct and lead to a much faster key distribution system than single-photon QKD systems. These are also relevant results that allow for a renewal of bit-to-bit encryption protocols.

The paper is organised as follows. In the next section, we introduce basic properties of coherent states of light. In Section~\ref{sec:protocol}, we present the protocol. In the subsequent Section~\ref{sec:security}, we analyse the protocol's security. Finally, in the last section, we present conclusions and some possible future lines of research.

\section{Coherent states of light}
\label{sec:coherent_states}

For simplicity, we consider only single-mode states of light, given by the annihilation operator $\hat a$. The ground state $\ket 0$ of the Hamiltonian $\hat H = \hbar\omega(\hat a^\dag \hat a + \frac 1 2)$, also called the {\em vacuum}, determines the orthonormal basis $\{ \ket n = \frac{1}{\sqrt{n!}} \hat a^n \ket 0 | n\in\mathbb N_0 \}$, called the {\em number basis}.  Coherent states are given by the following expression:
\begin{equation}
\label{coherent}
\ket\alpha = e^{-\frac{|\alpha|^2}{2}} \sum_{n=0}^\infty \frac{\alpha^n}{\sqrt{n!}} \ket n,
\end{equation}
where $\alpha = e^{i\varphi} |\alpha| \in \mathbb C \setminus \{ 0 \}$. Coherent states saturate Heisenberg relations for the position and the momentum operators $\hat x = \sqrt{\frac{m\omega\hbar}{2}}(\hat a^\dag + \hat a)$ and $\hat p = i\sqrt{\frac{\hbar}{2m\omega}}(\hat a^\dag - \hat a)$, i.e., $\Delta\hat x \Delta\hat p = \hbar/2$. Moreover, the expectation values $\langle\hat x\rangle$ and $\langle\hat p\rangle$ for coherent states obey position and momentum equations of motion of a classical harmonic oscillator. Therefore, they are considered to be the most ``classical'' quantum states (for more details on coherent states, see for example a comprehensive review~\cite{zha:90}).

One can introduce the so-called number operator $\hat n = \hat a^\dag \hat a = \sum_{n=0}^\infty n \ket n \bra n$ which counts the photon-number. For coherent states $\ket\alpha$ the average photon-number, i.e., its intensity, is $\langle \hat n \rangle = |\alpha|^2$. Moreover, the number operator is generator for the phase-rotation operator $\hat R (\varphi) = e^{- i \hat n \varphi}$. Given light-pulse intensity $\langle \hat n \rangle = |\alpha|^2$ and choosing a ``reference value'' $\alpha$, one can define states:
\begin{equation}
\label{modulated_state}
\ket{\Psi(\varphi)} = \hat R (\varphi) \ket\alpha = \ket{e^{- i \varphi} \alpha}.
\end{equation}
For each two angles $\varphi \neq \varphi^\prime$, the corresponding states $\ket{\Psi(\varphi)}$ and $\ket{\Psi(\varphi^\prime)}$ have a non-zero overlap. In the limit $\Delta\varphi  \rightarrow 0$, we have:
$$\begin{array}{rcl}
|\bracket{\Psi(\varphi)}{\Psi(\varphi^\prime)}|^2 
    &=& \exp [-4|\alpha|^2 (\sin \frac{\Delta\varphi}{2})^2] \\[2mm]
    &\xrightarrow{\Delta\varphi \rightarrow 0}& \exp [-|\alpha|^2 \Delta\varphi^2]\\[2mm]
    &=& \exp [- \frac{(\varphi - \varphi^\prime)^2}{2\sigma^2_\alpha}],
\end{array}$$
with $\sigma^2_{\alpha} = 1/(2|\alpha|^2) = 1/(2\langle \hat n\rangle)$. On the other hand, for $|\Delta\varphi| = |\varphi - \varphi^\prime | = \pi$, even for modest values of the average number of photons,
%(i.e., $\langle \hat n \rangle = |\alpha|^2 \gg 1$)
the two states become quasi-orthogonal:
$$\begin{array}{rcl}
|\bracket{\Psi(\varphi)}{\Psi(\varphi^\prime)}|^2 
    &=& \exp [-2|\alpha|^2 (1-\cos \Delta \varphi)] \\[2mm]
    &\xrightarrow{|\Delta\varphi| = \pi}& \exp [-4|\alpha|^2 ]\\[2mm]
    &\xrightarrow{|\alpha|^2 \gg 1}& 0.
\end{array}$$
Thus, each two states $\ket{\Psi(\varphi)}$ and $\ket{\Psi(\varphi + \pi)}$ form a (quasi-orthogonal) basis. In our protocol, we will consider $2M$ discrete bases $\mathcal{B}_k = \{ \ket{\Psi(\varphi_k)}, \ket{\Psi(\varphi_k^{\bot})}\}$, given by the angles $\varphi_k = k\frac \pi M$ and $\varphi_k^\bot = k\frac{\pi}{M} + \pi$, with $k = 0, 1, \dots 2M-1$ and $M \in \mathbb N$. As quantum states are used to encode bit values 0 and 1, we will consider bases  $\mathcal{B}_k$ to be ordered, such that the first vector $\ket{\Psi(\varphi_k)}$ encodes bit value 0, and the second, $\ket{\Psi(\varphi_k^\bot)}$, encodes bit value 1. Note that for each $k$, there exists the corresponding $\tilde k = k+M \pmod{2M}$, such that the two bases $\mathcal{B}_k$ and $\mathcal{B}_{\tilde k}$ consist of the same two states, encoding the opposite bit values. The ``reference'' basis $\mathcal{B}_0$ we will call the {\em computational basis}, with $\ket 0 = \ket{\Psi(0)} = \ket \alpha$ and $\ket 1 = \ket{\Psi(\pi)} = \ket{e^{-i\pi}\alpha}$. Finally, we will use the term ``measurement in a (computational) basis'' in a sense of ``performing the optimal state discrimination between the two basis states ($\ket{\Psi(0)}$ and $\ket{\Psi(1)}$)''. According to the Helstrom bound~\cite{hel:76}, the minimum probability of making an error in ambiguously inferring between two pure states $\ket\psi$ and $\ket\phi$ is $P_e = \frac{1}{2}(1 - \sqrt{1 - |\langle\psi | \phi\rangle |^2})$, which for two quasi-orthogonal states from $\mathcal{B}_k$ gives effectively perfect discrimination. Optimal discrimination between quasi-orthogonal coherent states in experiments is customary and can be efficiently performed in the lab (see for example the M-ry phase encoding procedure or amplitude encoding scheme~\cite{cor:bar:lia:que:kum:03,cor:etal:05,hir:etal:09}).

\section{The protocol}
\label{sec:protocol}

We combine the ideas of secure key distribution with continuous variables presented in~\cite{bar:03,bar:gra:15} with the secure message transmission scheme introduced in~\cite{nik:08}.
From a high level point of view, the latter scheme is as follows:
Alice prepares a quantum system consisting of qubits,
whose orientations are chosen at random but known to Alice,
and sends it to Bob. 
He encodes his bits by doing nothing for a 0,
and applying a phase flip (rotation of $\pi$) for a 1,
and then returns the system to Alice.
Since she prepared the system
she knows how to measure each qubit
and can retrieve the bit encoded.
Our protocol follows the same overall idea,
but instead of using qubits (two-dimensional quantum systems),
we use coherent states of light pulses to encode the bit values.

The functionality presented in~\cite{nik:08}
is that of secure message transmission
from Alice to Bob.
However, we prefer to cast our techniques
as a key distribution scheme, with
the obvious option of using the resulting key $r$
as input to an additional One-Time Pad encryption round.
Note that this modification does not change the underlying
physics; it merely adds an additional round
over the classical channel.
And it offers more flexibility
since the resulting key can be used
to send a message from Alice to Bob, from Bob to Alice,
or for message authentication.
The reason for preferring key distribution over message transmission 
is that if the input for the protocol is a random string,
then several kinds of post-processing techniques
(such as privacy amplification) are permitted,
whereas if the input is a plaintext message
then post-processing must not degrade the message.
Also, an abort when information has leaked will be too late
in the latter case.

Before giving more details about the quantum part,
observe that Alice and Bob also dispose of a classical communication channel,
more precisely a {\em public authenticated channel}.
By this we mean a broadcast channel which does not provide privacy 
(i.e., anyone can listen to a conversation), 
but does provide message and source authentication -- 
Eve cannot tamper with a message sent by Alice or Bob. Note that the existence of such a channel is 
a standard assumption in most quantum key distribution,
including BB84~\cite{ben:bra:84}.

As explained above,
the quantum part of the protocol consists of three steps:
(1) Alice sends to Bob $K$ distinguishable pulses of coherent light, 
such that each pulse $j=1 \dots K$ encodes the bit value $0$ in the $\mathcal{B}_{k_j}$ basis,
 i.e., is in state $\ket{\Psi(\varphi_{k_j})}_j$.
[The pulses are distinguishable by the time or the place of emission (depending on wether they are produced sequentially in time by a single laser, or in parallel, by a number of different lasers), denoted by the label $j$ of the kets (i.e., each light pulse is from a different Hilbert space $\mathcal{H}_j$).]
  This way, Alice produces a  multi-photon quantum system whose state $\ket{\psi_k} = \otimes_{j=1}^K \ket{\Psi(\varphi_{k_j})}_j$ is given by a tensor product of $K$ coherent pulses.
(2) Each pulse $\ket{\Psi(\varphi_{k_j})}_j$ will be used to encode a single bit value of the key $r$, by rotating its phase by $\pi$, in case $r_j=1$, and doing nothing in case $r_j=0$. 
(3) Alice rotates back each pulse to its computational basis and reads bit string $r$. 
In other words, our protocol can be interpreted as a quantum one-time pad generalisation in which, instead of two distinguishable classical/orthogonal/basis states, Alice chooses between $2M$ partially distinguishable quantum bases.

Now, the simplest way for a malicious Eve to eavesdrop the communication is 
a full man-in-the-middle attack: she intercepts Alice's
quantum state $\ket{\psi_k}$ and keeps it stored in a stable quantum memory (say, a delay device such as an optical fibre pool),
while sending her own state $\ket{\psi_e}$ to Bob,
which the latter will use to encode his key $r$. 
Now Eve, upon intercepting Bob's state $\ket{\psi_e(r)}$ encoded with her own $e$,
can easily decode it to learn $r$, and forward it to Alice encoded in the state $\ket{\psi_k(r)}$.

The above attack can be easily avoided by a simple verification technique:
when she sends the quantum state $\ket{\psi_k}$,
Alice also provides Bob, through the authenticated channel,
with a certain number (say, $K/2$)
of randomly chosen bases $k_{j}$ of $k$, thus allowing Bob to check if the partial
states $\hat\rho_{k_{j}} = \Tr_{k \setminus k_{j}} \ket\psi\bra\psi$
of the $j$-th light pulse are indeed the expected pure states
$\hat R (\varphi_{k_{j}}) \ket\alpha$.
This explains the inclusion of Steps 2b and 2c in Protocol 1 below.
This verification technique should also be
applied at the end of Step 2, with the roles reversed, 
to avoid Eve tampering with
the state $\ket{\psi_k(r)}$ sent from Bob to Alice.
This corresponds to Steps 2f and 3a below.

This verification technique is fairly standard in quantum cryptography (used for example in famous quantum key distribution schemes). It severely restricts Eve's class of attacks. For example, she cannot split coherent-light pulses, keeping one part with her, as this change of state would be easy to spot by measuring the average photon number). 
So Bob and Alice must receive the intended states $\ket{\psi_k}$ and $\ket{\psi_k(r)}$, at the beginning of Steps 2 and 3, respectively. 
Eve can only correlate her ancilla systems 
(on which she can subsequently perform measurements) with $\ket{\psi_k}$ and $\ket{\psi_k(r)}$, respectively. 

This leads to the following protocol.

\begin{protocol}[QKD by phase flip]\label{prot:PK} \em
\ \\[2mm]
{\em Setup:}
\noindent
\begin{itemize}
\item $\langle \hat n\rangle=|\alpha|^2$ : expected photon number\ \\[-3mm]
\item $2M$ : number of possible bases\ \\[-3mm]
\item $ K $ : initial number of pulses\ \\[-3mm]
\item $k=(k_1,\dots,k_K)$, where each $k_j\in\{0,\dots, 2M-1\}$ : Alice's choice of bases\ \\[-3mm]
\item $r=(r_1,\dots,r_K')$, where each $r_j \in \{ 0,1 \}$ : Bob's choice of key bits
\end{itemize}

\noindent
{\em{\bf Step 1.} Preparation of the quantum state} \
\begin{itemize}
    \item[(a)] For $1 \leq j \leq K$, Alice chooses uniformly at random $k_j~\in~\{0,\dots, 2M-1\}$, forming his choice of bases $k=(k_1,\dots,k_K)$.\ \\[-3mm]
    \item[(b)] Alice prepares the corresponding quantum state
    \begin{equation}
    \ket{\psi_k}    = 
        \bigotimes_{j=1}^K \left[ \hat R (\varphi_{k_j}) \ket\alpha_j \right]= \bigotimes_{j=1}^K \ket{e^{-i\varphi_{k_j}}\alpha}_j
    \end{equation}
    and sends it to Bob.
\end{itemize}

\noindent
{\em {\bf Step 2.} Encoding the random key $r$} \
    \begin{itemize}
    \item[(a)]
    Bob receives $\ket{\psi_k}$ and informs Alice of that fact.\ \\[-3mm]
\item[(b)]
    Alice chooses a random bit string $v$ of size $K$ and weight $K'=K/2$ (i.e., a string with equal number of zeros and ones).
    Then she computes $k'$,
    where $k'_j = k_j$ if $v_j = 0$, and $ k'_j = \square $ otherwise,
     and sends $v$ and $k'$ to Bob over
    the authenticated channel. Here, $\square$ represents a default value different from any possible value of $k_j$, indicating that the pulses for which $v_j = 1$ will not be used in the Bob's verification procedure (the following Step 2 (c)).\ \\[-3mm]
        \item[(c)] 
        Bob receives  $v, k'$
    and verifies that for $j$ with $v_j = 0$,
     $\hat\rho_{k'_{j}} $ equals the pure state
$\hat R (\varphi_{k'_{j}}) \ket\alpha_j$. \ \\[-3mm]
    \item [(d)]
    Let $\ket{\psi'_k}$ denote the quantum state received by Bob
    in which positions with $v_i = 0$ have been traced out.
    Bob generates a random string $r$ of size $K/2$, the key,   
    and encrypts it as follows:
    \begin{equation}
    \ket{\psi'_k(r)} =
    \left[  \bigotimes_{j=1}^{K'} \hat R(r_j \pi) \right] \ket{\psi_k}.
    \end{equation}
    
    \item[(e)] 
    Bob sends $\ket{\psi'_k(r)}$ to Alice.\ \\[-3mm]
    \item[(f)]
    Bob chooses a random bit string $w$ of size $K' = K/2$ and weight $K''=K/4$. Then he computes $r'$, where $r'_j = r_j$ if $w_j = 0$, and $ r'_j = \square $ otherwise, and sends $w$ and $r'$ to Alice over the authenticated channel.
    He computes the final key $r$ which is obtained from $r$    
    by concatenating all the bit positions
    $r_i$ for which $w_i=1$.
    
    \end{itemize}

\noindent
    {\em {\bf Step 3.} Decoding the random key $r$} \
    \begin{enumerate}[(a)]
    \item
    Alice receives $\ket{\psi'_k}$
    and uses her choice of bases $k$ to verify that for $j$ with $w_j = 0$,
     $\hat\rho_{k_{j}} $ equals the pure state
$\hat R (\varphi_{k_{j}} \!\! +  r_j\pi) \ket\alpha_j$. \ \\[-3mm]
    \item Then she uses the remaining positions, i.e.\ with $w_j=1$, to determine $r$ of size $K''$ encoded in quantum states of the computational basis $\mathcal{B}_0$:
    \begin{eqnarray}
    \hspace*{-15mm}\ket{\psi''(r)} &=&  
     \left[ \bigotimes_{j=1}^{K''} \hat R( -\varphi_{k_j}) \right] \ket{\psi_k(r)} \nonumber \\ 
    &=& \ds  
    \bigotimes_{j=1}^{K''} \ket{e^{-ir_j \pi}\alpha}_j\\
    &=& \ds  
    \bigotimes_{j=1}^{K''} \ket{r_j}_j. \nonumber
    \end{eqnarray}
    \end{enumerate}
\end{protocol}

Note that, in order to perform state verification in Step 2c, Bob has to store $\ket{\psi_k}$ in a stable quantum memory, while waiting for Alice to send him the needed classical information, during Step 2b. The existence of long-term stable quantum memories is currently still a matter of considerable technological limitations, which might seem to undermine the security of current implementation of our protocol. However, while indeed the lack of quantum memories prevents the implementation of the verification procedure, it also prevents Eve from performing the man-in-the-middle attack, the very reason for the need for verification. In other words, as long as stable quantum memories are out of the reach of the technology, Eve would not be able to perform the attack which would require the verification procedure, and consequently the protocol security would not be compromised by this fact. It is an interesting question, though, to analyse the case of Eve and Bob having realistic, noisy memories, but of a different quality -- say, Eve is a wealthy corporation/agency that wants to breach the privacy of ordinary everyday consumers who cannot afford the expensive cutting-edge technology. While indeed interesting and relevant, such analysis exceeds the scope of this paper (for the analysis of the effects of  realistic noisy memories on the security on a two-state quantum bit-commitment protocol, see~\cite{lou:14}).

\section{Security of the protocol}
\label{sec:security}

In the following, we analyse in more detail the protocol's security against attacks in which Eve intercepts pulses sent by Alice, performs measurements on them and forwards the pulses to Bob, or just entangles her ancillas with the pulses and measures the pulses returned by Bob after Step 2, as well as against the beam splitting attack.  First, we show the security of $k$, Alice's choice of bases: upon intercepting the state $\ket{\psi_k}$, Eve cannot learn the chosen bases $k$, and consequently she cannot decode the key $r$. Note that from the point of view of Eve, who does not know $k$, the mixed state $\hat\rho_B$ of an array of $K$ pulses of coherent light sent by Alice is:
\begin{eqnarray}
\hat\rho _B & = & \frac{1}{(2M)^K} \sum_{k_1,\dots k_K = 0}^{2M-1} \left[ \bigotimes_{j=1}^K \ket{\Psi(\varphi_{k_j})}\bra{\Psi(\varphi_{k_j})} \right] \nonumber \\ & = & (\hat\rho)^{\otimes K},
\end{eqnarray}
where $\hat\rho = \frac{1}{2M} \sum_{k= 0}^{2M-1} \ket{\Psi(\varphi_{k})}\bra{\Psi(\varphi_{k})}$. The summation over $k_j$ implies the lack of knowledge of Eve on the basis used among the possible bases. Note that both $\hat\rho_B$ and $\hat\rho$ are implicitly functions of $M$ and $\langle \hat n \rangle = |\alpha|^2$. The Holevo Theorem says that upon performing an arbitrary POVM on $\ket{\psi_k}$, the mutual information $I(k:e)$ between $k$ and Eve's inference $e$ is bounded by the von Neumann entropy $S(\hat\rho_B)$ of the state $\hat\rho_B$:
\begin{equation}
I(k:e) \leq S(\hat\rho_B) = K \! \cdot \! S(\hat\rho).
\end{equation} 
On the other hand, the Shannon entropy $H(k)$ of $k$ is: 
\begin{equation}
H(k) = K \! \cdot \! \log M.
\end{equation}
In other words, in order not to allow Eve to learn  more than a negligible part of $k$, the following equation has to be satisfied:
\begin{equation}
\label{criterion}
S(\hat\rho) \ll \log M.
\end{equation}
 
From equation~\eqref{coherent} for coherent states, one can obtain the following expression for $\hat\rho$:
\begin{eqnarray}
\hat\rho & = & \frac{1}{2M} \sum_{k= 0}^{2M-1} \ket{\Psi(\varphi_{k})}\bra{\Psi(\varphi_{k})}  \\ & = & \frac{e^{-|\alpha|^2}}{2M} \sum_{n,n'=0}^{+\infty} \frac{(\alpha)^n(\alpha^\ast)^{n'}}{\sqrt{n!n'!}} J(n,n';M) \ket n \bra {n'}, \nonumber  
\end{eqnarray}
where $J(n,n';M) = \sum_{k=0}^{2M-1} e^{ik(n'-n)\frac{\pi}{2M}} = \sum_{k=0}^{2M-1} q^k$ and $q = e^{i(n'-n)\frac{\pi}{2M}}$. For $q\neq 1$ (i.e., $n\neq n'$), we have:
\begin{equation}
    \ds J(n,n';M) =
        \ds
            \left\{\!\!\!
                \begin{array}{cl}
                    \ds 2M, & \!\!\!\!\! \mbox{ if } n' - n = 0\\
                    \ds 0, & \!\!\!\!\! \mbox{ if } n'-n=2l, l\in \mathbb{Z} \setminus \{ 0 \} \\
                     \ds - \frac{2}{e^{i\frac{2l+1}{2M}\pi} - 1},
                                                & \!\!\!\!\! \mbox{ if } n'-n=2l +1, l\in \mathbb{Z} \setminus \{ 0 \}.
                \end{array}
            \right.
\end{equation}

Finally, one gets
\begin{equation}
\hat\rho = \sum_{n\in\mathbb{N}_0} \rho_{n,n} \ket n \bra n + \sum_{n\in\mathbb{N}_0} \rho_{n,n+2l+1} \ket n \bra{n+2l+1},
\end{equation}

with the matrix elements given by ($\alpha = |\alpha|e^{i\theta}$):
\begin{eqnarray}
\label{matrix-elements}
\rho_{n,n} & = & \frac{e^{-|\alpha|^2} |\alpha|^{2n}}{n!}  \\ \rho_{n,n+2l+1} & = & \frac{e^{-|\alpha|^2} |\alpha|^{2(n+l)+1}}{M\sqrt{n!(n+2l+1)!}} \cdot \frac{e^{-i(2l+1)\theta}}{e^{(l+\frac{1}{2})\frac{\pi}{m}} - 1}. \nonumber  
\end{eqnarray}

\begin{figure}[!hb]
\centering
\ \\[-0.2cm]
\includegraphics[width=8.5cm,height=6.0cm,angle=0]{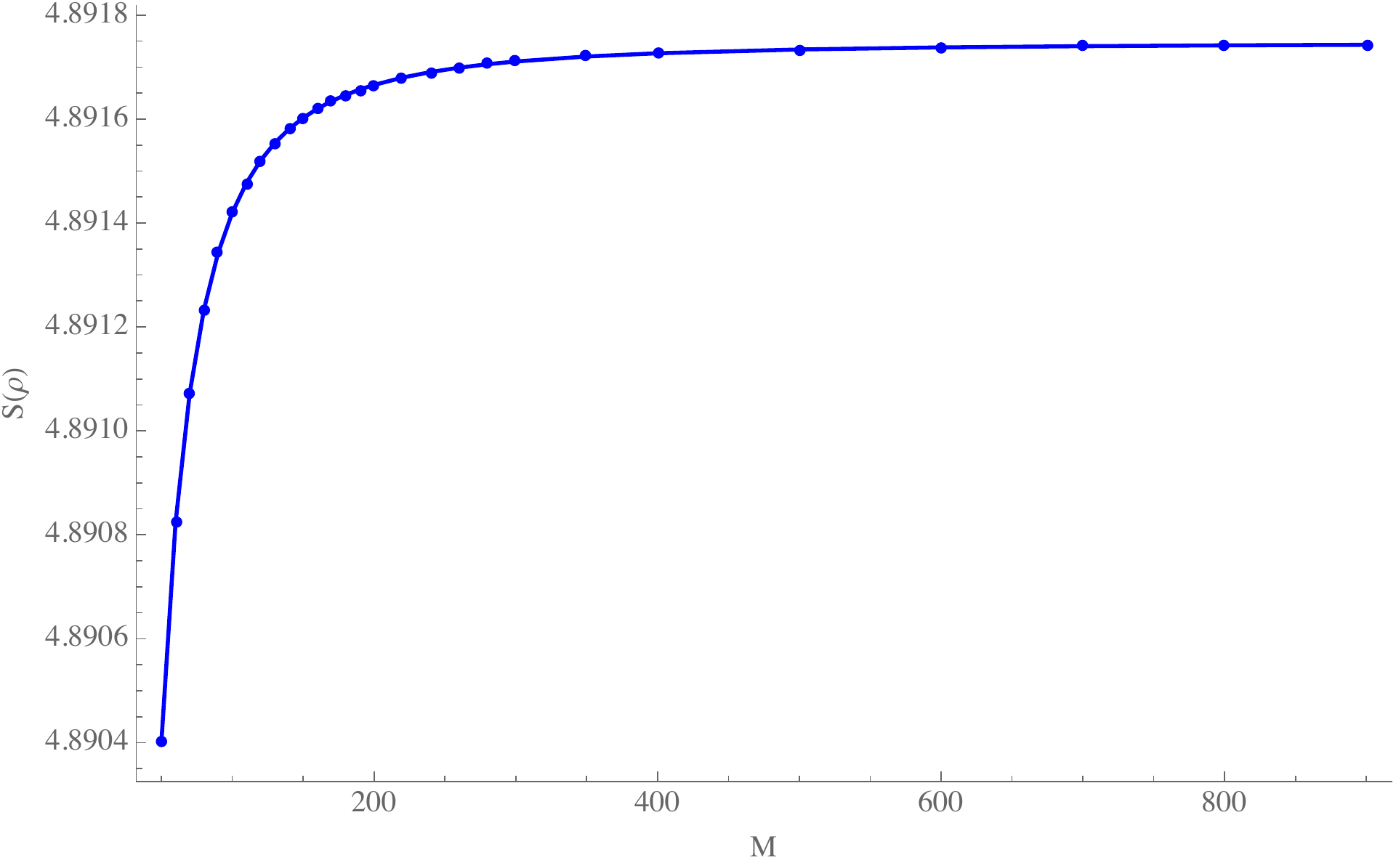}
\ \\[-0.2cm]
\caption{(color online) The plot of $S(\hat\rho)$ as a function of $M$ for $\langle \hat n \rangle = |\alpha|^2 = 200$.} \label{S(R)_20}
\end{figure}

\begin{figure}[!hb]
\centering
\ \\[-0.2cm]
\includegraphics[width=8.5cm,height=6.0cm,angle=0]{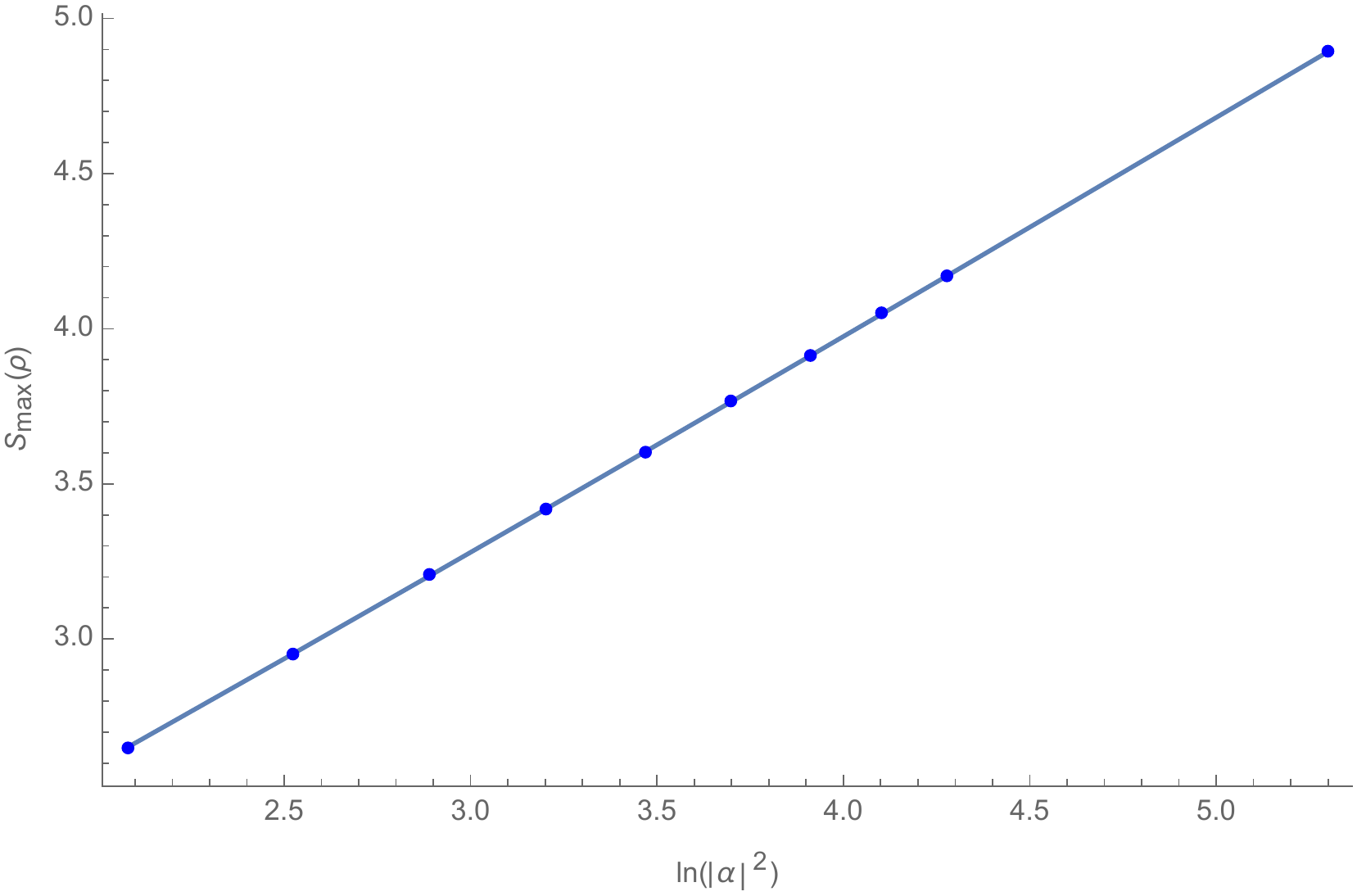}
\ \\[-0.2cm]
\caption{(color online) The logarithmic plot of  $S_{max}(\hat\rho)$ as a function of $\ln (|\alpha |^2)$, for $\langle \hat n \rangle = |\alpha|^2 =  8, \dots , 200$.} \label{S_m(alpha)}
\end{figure}

Numerical results for $S(\hat\rho)$ confirm that the above criterion~\eqref{criterion} is satisfied even for modest values on photons per pulse. On Figure~\ref{S(R)_20} we present $S(\hat\rho)$ as a function of $M$, for $\langle \hat n \rangle = |\alpha|^2 = 200$ (note that due to the symmetry, the von Neumann entropy is not a function of the phase of the ``reference value'' $\alpha$). We see that after a steep increase, the curve reaches a plateau $S_{max}(\hat\rho)$, showing that for big enough $M$ the above criterion~\eqref{criterion} is satisfied. On Figure~\ref{S_m(alpha)} we plot  $S_{max}(\hat\rho)$ for $\langle \hat n \rangle = |\alpha |^2 = 8, \dots ,200$ showing that the security criterion~\eqref{criterion} is satisfied for a wide range of photon-numbers.

%_________________________FIGURE

%__________________________________

%_________________________FIGURE

%__________________________________

Moreover, one can give an upper bound to $S_{max}(\hat\rho)$ confirming the above plot from Figure~\ref{S_m(alpha)}. Since any non-selective measurement increases the entropy, one has:
\begin{equation}
\label{bound}
S(\hat\rho) \leq S \left(\sum_n \ket n \bra n \hat\rho \ket n \bra n \right) = H(\{ p_n^{(\hat n )} \}),
\end{equation}
where $p_n^{(\hat n )} = \bra n \hat\rho \ket n$ is the probability to find $n$ photons in the pulse. The results of the measurement of the number operator $\hat n$ obey the Poisson distribution (the diagonal elements of $\hat\rho$, see~\eqref{matrix-elements}), for which the Shannon entropy's leading term in the large-$|\alpha|$ asymptotic expansion is precisely of the order of $\ln |\alpha|$. Thus, in order to satisfy the security criterion~\eqref{criterion}, one must have $\langle \hat n \rangle = |\alpha |^2 \ll M$.

One might consider an analogous to the beam-splitting attack used in the case of single-photon key-distribution cryptography: Eve can try to split the coherent pulse sent from Alice, send one of its parts to Bob, and keep the rest with her. Unlike the single-photon case, where the signals are weak (precisely in order to minimise the probability of multi-photon emissions), in our case of genuine multi-photon coherent pulses, one can measure photon-number as well. Indeed, since the average photon number is given by $\langle \hat n \rangle = |\alpha |^2$, this is precisely what Bob does in Step 2 (c): ``verifies that for $j$ with $v_j = 0$, $\hat\rho_{k'_{j}} $ equals the pure state $\hat R (\varphi_{k'_{j}}) \ket\alpha_j$'' (obviously, verifying that the state is given by a complex number $e^{-i\varphi_{k'_j}}\alpha_j$ goes beyond just checking its absolute value $|\alpha_j|^2$).

In realistic implementations, the photon losses during the emission, transmission and detection, as well as due to imperfect detectors, lead to an effective decrease of the mean photon-number $|\alpha|^2$. Nevertheless, the users of the devices, Alice and Bob, do have the knowledge of the original laser intensity and overall efficiency of the fibre and detectors used, provided by the lab/company that have assembled and maintain the network (much as any user is provided by the essential specifications of a product needed for it to be properly used). This way, Bob can in advance anticipate the losses of an untempered network, and detect eavesdropping through additional losses. 

One of the common attacks widely considered regarding quantum key distribution protocols is the so-called collective attack. Such attacks, designed for the BB84-like key distribution, is not really applicable in our case. In BB84 (and similar protocols), upon exchanging quantum systems and performing the measurements, Alice and Bob exchange classical information over the public network that allows them to extract the key. The collective attacks are the attacks in which Eve intercepts quantum communication from Alice to Bob, entangles her ancilla qubits to those sent by Alice, resends the intercepted qubits to Bob, and waits to perform the measurement(s) on her ancillas only upon learning the subsequently exchanged classical information. But in our protocol, no classical information is exchanged between Alice and Bob regarding the pulses used for extracting the key, apart from the trivial information on the success of the final measurement performed by Alice, so this type of attack is not applicable in our case.

Eve could in principle use more sophisticated techniques to, upon intercepting the encoded state $\ket{\psi_k (r)}$, ``directly'' learn the key $r$. For instance, she could initially entangle her ancilla with the pulses sent by Alice in such a way to preserve the classical correlations between the bases $k$ and the quantum states $\ket{\psi_k}$, thus succeeding to pass the protocol's verification procedure, and finally perform a joint coherent measurement on the pulses returned by Bob to Alice. As noted at the end of the previous section, showing the protocol's security against general coherent attacks under realistic effects of noise and measurement errors exceeds the scope of this paper. Following the ideas of device-independent quantum cryptography~\cite{vaz:vid:14}, one could introduce an additional verification based on the violation of a Bell-like inequality and on entanglement monogamy, which could in principle allow Alice and Bob to detect arbitrary weak tampering by Eve. Nevertheless, mesoscopic continuous variable entangled states are far beyond today's technology.

Furthermore, one should keep in mind that the proposed system can work in long haul channels where several amplification stages are used (see Section V of~\cite{bar:05}). Tapping techniques that could be used by Eve based on single photons cannot work through optical amplifiers, due to the non-cloning theorem. Finally, the mesoscopic signals also demand some bandwidth separation from other channels whenever a same optical fiber is used to avoid signal ``talking" from the intense modes with the mesoscopic one. 

\section{Conclusions}
\label{sec:conclusion}

In this paper, we have presented a quantum key distribution protocol that uses coherent states of light to encode single-bit values, and analysed its security. The protocol is based on the interplay between the quantum noise and the information acquired by Eve (given by the Holevo's theorem): whenever the laser signals are strong enough so that the noise over signal ratio $\langle (\Delta \hat n)^2 \rangle / \langle \hat n \rangle^2$ goes to zero, the signal can be seen as a classical signal and can be perfectly copied. In the opposite case, if the shot noise is high enough, no measurement will produce identical results on similarly prepared signals. This is the physical protection behind this communication -- the attacker cannot distinguish between the signals sent from Bob to Alice in the public communication stage. This establishes the basic condition for the application of Holevo's theorem.

Moreover, while the protocol introduced in~\cite{bar:03,bar:gra:15} requires for certain amount of a pre-shared secret key, our protocol does not (a feature shared with the protocol presented in~\cite{nik:08}). However, unlike the protocol from~\cite{nik:08}, which is based on single-qubit rotations, our protocol is not constrained by the need of slow and expensive detectors, characteristic for applications that encode bit values in single-photon states. Finally, the ``Holevo argument'' is rather non-trivial in our case, as single bit values are encoded in states from an infinitely-dimensional Hilbert space (and not in 2-dimensional qubit states, as is the case of~\cite{nik:08}). We confirmed numerically that for a huge range of the average photon-number per pulse, the maximal amount of information that can be transmitted by a single pulse of coherent light, as a function of $M$, saturates to a finite value. Moreover, by giving the upper bound to the von Neumann entropy $S_{max}(\hat\rho)$, we confirmed the numerically observed logarithmic behaviour of its dependance on $|\alpha |$, thereby ensuring the protocol's security, by choosing sufficiently large $M \gg \langle \hat n \rangle = |\alpha |^2$. We also showed that the protocol is secure against a beam-splitting attacks, and that the analog of the collective attack is not applicable to the case of our protocol. %Our protocol can be modified to a semi-quantum protocol, providing a cheap solution for everyday consumers, who do not require fully sophisticated quantum equipment.

It should be emphasized that the presented protocol is a fast protocol due to the light intensities involved, the fast detectors utilized (telecomm type) and the lower associated cost. The presented protocol is not a single photon based protocol neither depend on short distances under losses to operate as usual in the BB84 kind of protocols. Also the objective is not to present any substitutive for BB84 or the single photon family but just to present a practical and fast system for many applications that demand long range, lower cost and fast speeds.
 
The main direction of the future work would be to analyse quantitatively the protocol's security level as a function of security parameters $M$ and $\langle \hat n \rangle = |\alpha |^2$ against concrete attacks (single-qubit measurements only, etc.). One can also study other cryptographic protocols with continuous variables based on the public key encryption scheme used in this article. For example, it is possible to straightforwardly use coherent states instead of two-dimensional states of qubits to achieve more robust oblivious transfer protocol presented in~\cite{rod:mat:pau:sou:14}.

\vspace{0.4cm}

\section*{Acknowledgments}

\noindent PM and NP thank Antony Laverrier for fruitful discussions and for bringing us the attention to the recent work~\cite{ott:man:pir:15,zhu:zha:dov:won:sha:16}.

This work was partially supported, by SQIG at Intituto de Telecomunica\c{c}\~oes, and the CV-Quantum and QbigD internal projects at IT funded by FCT PEst-OE/EEI/LA0008/2013 and UID/EEA/50008/2013.

%\section*{References}
\bibliography{CV_QKD}

\end{document}